\documentclass[fleqn,usenatbib]{mnras}

\usepackage{newtxtext,newtxmath}

\usepackage[T1]{fontenc}

\DeclareRobustCommand{\VAN}[3]{#2}
\let\VANthebibliography\thebibliography
\def\thebibliography{\DeclareRobustCommand{\VAN}[3]{##3}\VANthebibliography}

\usepackage{graphicx}	
\usepackage{amsmath}	

\newcommand{\hii}{\ion{H}{II}}
\newcommand{\kms}{km~s$^{-1}$} 
 
\title[Ethynyl around S255 and S257]{Ethynyl around the \hii{} regions S255 and S257\thanks{This work is based on observations carried out under the project 027-18 with the IRAM~30~m Telescope. IRAM is supported by INSU/CNRS (France), MPG (Germany), and IGN (Spain).}}

\author[A. I. Buslaeva et al.]{
Anna I. Buslaeva,$^{1}$
Maria. S. Kirsanova,$^{1}$\thanks{E-mail: kirsanova@inasan.ru (MSK)}
Anna F. Punanova$^{2}$\\
$^{1}$Institute of Astronomy, Russian Academy of Sciences, 119017, 48 Pyatnitskaya Str., Moscow, Russia\\
$^{2}$Institute of Natural Sciences and Mathematics, Ural Federal University, 19 Mira Str., 620075 Ekaterinburg, Russia\\
}

\date{Accepted January 29, 2021. Received July 5, 2020}

\pubyear{2019}
\begin{document}
\label{firstpage}
\pagerange{\pageref{firstpage}--\pageref{lastpage}}
\maketitle

\begin{abstract}
We present the results of the ethynyl (C$_2$H) emission line observations towards the \hii{} regions S255 and S257 and the molecular cloud between them. Radial profiles of line brightness, column density, and abundance of C$_2$H are obtained. We show that the radial profile of the ethynyl abundance is almost flat towards the \hii{} regions and drops by a factor of two towards the molecular cloud. At the same time, we find that the abundance of ethynyl is at maximum towards the point sources in the molecular cloud -- the stars with emission lines or emitting in X-ray. The line profiles are consistent with the assumption that both \hii{} regions have front and back neutral walls that move relative to each other.
\end{abstract}




\section{Introduction}

Over the past 80 years, more and more new molecules have been discovered in the interstellar medium \citep{2013RvMP...85.1021T}. To date, more than 200 chemical species have been found in the dense and diffuse molecular clouds, apart to the atoms in different degrees of ionization in the \hii{} regions \citep[see, for example, the list of molecules at the website of the Cologne database for molecular spectroscopy CDMS\footnote{https://cdms.astro.uni-koeln.de/classic/molecules},][]{2016JMoSp.327...95E}. Complex organic molecules, such as methanol (CH$_3$OH), ethylene-glycol ((CH$_2$OH)$_2$), and many others, are formed by successive additions of individual atoms and radicals in the gas phase and on the surfaces of dust grains \citep{2013RvMP...85.1021T}. It is important that the so-called bottom-up chemistry requires reactions to occur on the surface of interstellar dust grains, which often serve as the catalysts for reactions. At the periphery of the \hii{} regions, in the photodissociation regions (PDR), where physical and chemical processes are caused by the absorption of the ultraviolet radiation with the energy below hydrogen ionization potential (UV, from 4 to 13.6~eV), the opposite process, namely, destruction of long carbon chains, so-called top-down chemistry occurs \citep{Berne2011, 2015A&A...577A.133B}. It is important to note that the main species (molecules, radicals) are different in the bottom-up and top-down chemistry~\citep{2013RvMP...85.1021T}. While the initial stages of the bottom-up chemistry are fairly well understood \citep[e.g., the chain of methanol formation from a CO molecule on the dust,][]{1982A&A...114..245T}, the starting point of the top-down chemistry and its main species are unknown, apart from the confirmed formation of fullerenes \citep{2014ApJ...797L..30Z} as products of photodissociation of polycyclic aromatic hydrocarbons (PAH). However, the top-down chemistry is known to be closely related to the effect of UV radiation upon PAH, hydrocarbon grains, and long carbon chains formed in the atmospheres of red giants \citep{2005A&A...435..885P, 2015ApJ...800L..33G, 2018IAUS..332....3V}.

Destruction of PAHs under the action of UV radiation  \citep{2013ARep...57..573P, 2019MNRAS.488..965M} can cause interesting features in the molecular composition of the gas in PDRs. For instance, \citet{2005A&A...435..885P} found high abundance of small hydrocarbons C$_2$H, c–C$_3$H$_2$, C$_4$H towards the illuminated edge of the Horsehead Nebula. The authors were unable to explain the formation of hydrocarbons at $A_{\rm V}=0.1$ using PDR model and suggested formation of small hydrocarbons as a result of the photo-destruction of PAHs. Later, \cite{2017A&A...605A..88L} simulated formation of hydrocarbons in the Horsehead Nebula, taking into account the presence of gas and dust in the PDR at the stage of a dark cloud before the beginning of ultraviolet irradiation by the $\sigma$~Ori star. They managed to explain the distribution of the molecular abundances everywhere, except for the illuminated edge of the Horsehead. \citet{2015ApJ...800L..33G} reported an excess of the l-C$_3$H$^+$ ion abundance towards the illuminated edge of the Horsehead as compared to the results of astrochemical modelling.

Interestingly, the Horsehead remains the only PDR where the abundance of small hydrocarbons has not been explained by the existing astrochemical models. In the Orion Bar PDR, close to the Orion nebula, high abundance of small hydrocarbons was explained by high-temperature chemical reactions involving excited hydrogen molecules \citep{2010ApJ...713..662A, 2015A&A...575A..82C}. Also, no discrepancies were found between the observed abundances of C$_2$H and the results of the astrochemical modelling in the M8 PDR by \citet{2019A&A...626A..28T}. \cite{Murga2020} showed that an effective formation of C$_2$H from acetylene molecules (C$_2$H$_2$) occurs as a result of photo-destruction of PAHs in the Orion Bar PDR, while in the Horsehead this process is practically insignificant in comparison to the formation of C$_2$H via gas-phase reactions starting from ionised carbon C$^+$. Experimental data on the efficiency of acetylene formation as a result of the photo-destruction of PAHs are contradictory. Some authors find that the products of PAH photo-destruction are C and H atoms, as well as H$_2$ molecules \citep[see, e.g.,][]{1994ApJ...420..307J, 2015ApJ...804L...7Z, 2019arXiv191203137J}.

In the infrared images obtained by space telescopes, \hii{} regions look like ring nebulae – projections of spherical shells or ring structures onto the plane of the sky~\citep{Deharveng_2005, 2010A&A...523A...6D}. Despite several thousand infrared nebulae have been catalogued to date, there are no more than ten well-studied PDRs around \hii{} regions~\citep{2006ApJ...649..759C, 2012MNRAS.424.2442S}. There is a lack of statistical data on the features of the size distribution of PAH particles and on the abundance of hydrocarbons in PDRs. Moreover, the profiles of the molecular abundances, obtained from the ionization source deep into the molecular cloud, are required for comparison with the results of theoretical modelling.

\begin{figure*}
\includegraphics[width=0.999\linewidth]{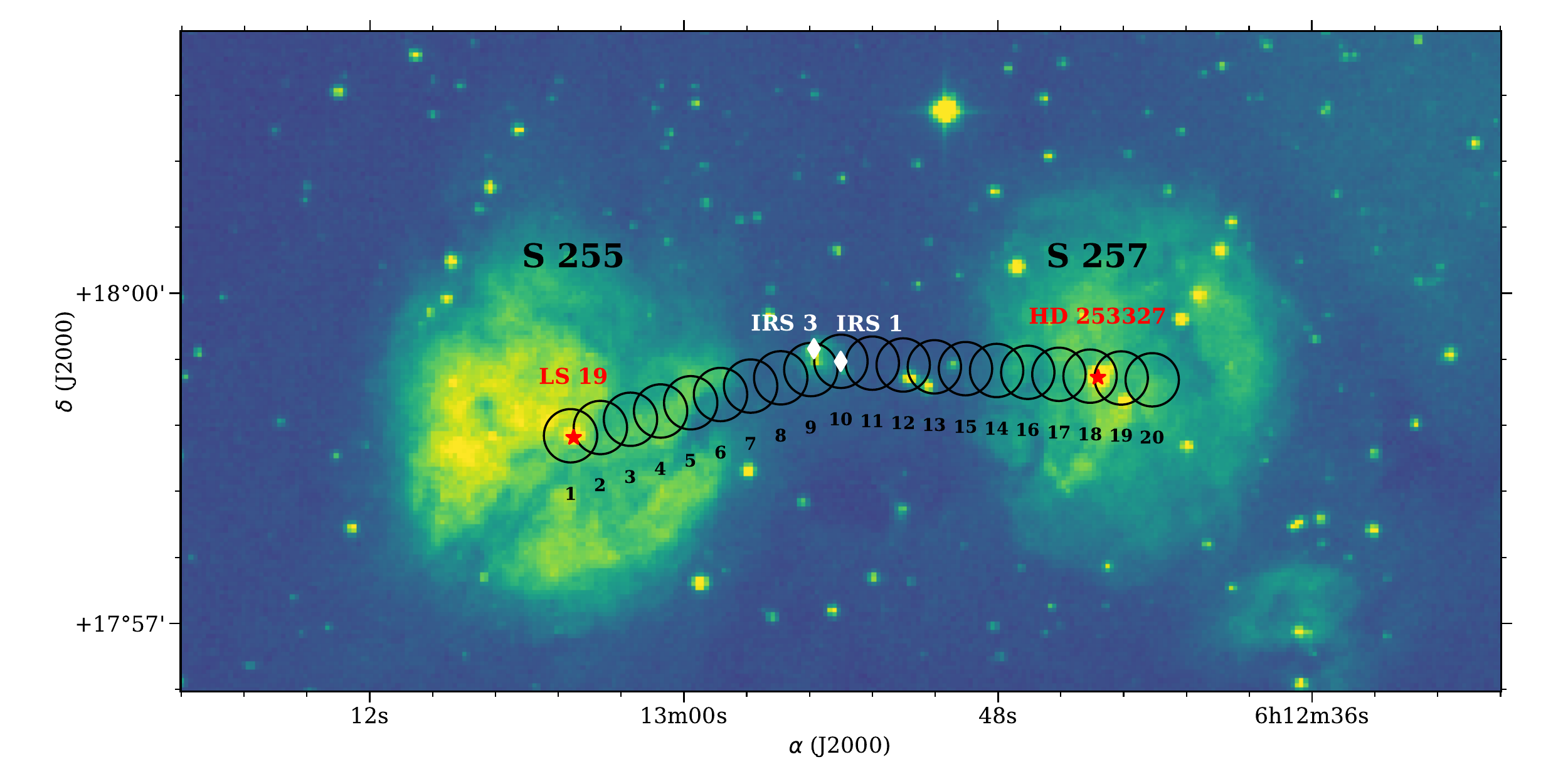}
\caption{Optical image of the \hii{} regions S255 and S257 from the DSS survey in the filter POSS2 Red. The observed positions are shown by empty circles whose sizes correspond to the telescope beam at the frequency 87~GHz (28\arcsec). Ionizing stars of the \hii{} regions are shown by red stars. IR-sources S255 IRS1 and S255 IRS3 are shown by white diamonds.}
\label{fig:DSS}
\end{figure*}

The aim of the present study is to obtain the radial profile of the C$_2$H abundances in the \hii{} regions Sh2-255 and Sh2-257 \citep[hereafter S255 and S257, see][]{1959ApJS....4..257S} in the complex S254-S258~\citep{2008ApJ...682..445C}. Entire complex is located at  $\approx 1.78$~kpc from the Sun~\citep{2007A&A...470..161R}. A molecular cloud is compressed between these two \hii{} regions, as found by \citet{2009AJ....138..975B, 2011ApJ...738..156O}, see Fig.~\ref{fig:DSS}. Spectral types of ionizing stars of these \hii{} regions are B0V ~\citep[LS~19 star in S255,][]{1979A&AS...38..197M} and B0.5V ~\citep[HD~253327 in S257][]{2011ApJ...738..156O}. The sizes of both \hii{} regions are about 1.3~pc. The neighbourhoods of S255 and S257 are active star-forming regions as was shown by~\cite{2008ApJ...682..445C, 2012ApJ...755..177Z, 2014MNRAS.439.3719C, 2020ApJ...889...43Z}. 

The details of our observations are presented in Sect.~\ref{section2}, and the {\it Herschel} Hi-GAL data used in the work are described in Sect.~\ref{herschel}. The spectral data analysis is described in Sect.~\ref{section3}, the results on the ethynil column density and abundance are presented in Sec.~\ref{section4}. We discuss the ethynil abundance profiles and the environmental context in Sect.~\ref{discussion} and give conclusions in Sect.~\ref{conclusions}.

\section{Observations}\label{section2}

In order to obtain the C$_2$H abundance profile in the S255 and S257 \hii{} regions, we made observations in several selected positions along the straight lines connecting the S255~IRS1 infrared source, which is located in the molecular cloud between the \hii{} regions (J2000 $\alpha$=06$^h$12$^m$54$^s$, $\delta$=17$^\circ$59$^\prime$23$^{\prime\prime}$), and the ionizing stars of each of the \hii{} regions: LS~19 (J2000 $\alpha$=06$^h$13$^m$04.2$^s$, $\delta$=17$^\circ$58$^\prime$41.5$^{\prime\prime}$) and HD~253327 (J2000 $\alpha$=06$^h$12$^m$44.2$^s$, $\delta$=17$^\circ$59$^\prime$14.3$^{\prime\prime}$), see Fig.\ref{fig:DSS}. Observations were done on 19--21 August 2018 with the IRAM 30~m telescope using the EMIR~090 receiver and the FTS~50 spectrograph in the full power mode. We made sure that the OFF-position (J2000 $\alpha$=06$^h$12$^m$54.0$^s$, $\delta$=18$^\circ$09$^\prime$23.0$^{\prime\prime}$) was free of the C$_2$H(1--0) emission down to the rms of 25~mK ($T_{\rm mb}$) using the frequency switching mode. The beam size at the frequency of 87.4~GHz was 28\arcsec, observations were carried out position-by-position with the step of 14\arcsec. Spectral resolution was 50~kHz, that at the frequency of 87.4~GHz corresponds to 0.17~\kms. Observations were carried out under acceptable weather conditions, the amount of precipitable water vapour pwv~=7–13~mm. The noise temperature of the system was at the level of 100~K. The intensity scale was converted to the main beam temperature ($T_{\rm mb}$) units taking into account the antenna efficiency $F_{\rm eff}=0.95$, $B_{\rm eff}=0.81$\footnote{\url{http://www.iram.es/IRAMES/mainWiki/Iram30mEfficiencies}}. The signal integration time at each point was 7.7~min, which made it possible to achieve the sensitivity of 0.05~K in units of $T_{\rm mb}$.
The processing and analysis of spectra (taking into account the antenna efficiency, subtraction of the baseline, approximation of the lines by the Gaussian profiles) were done using the CLASS/ GILDAS package\footnote{Continuum and Line Analysis Single-Dish Software \url{http://www.iram.fr/IRAMFR/GILDAS}}.

\section{Additional data}\label{herschel}

We used the data on the hydrogen column density along the line of sight ($N_{\rm H_2}$) and the dust temperature ($T_{\rm dust}$), obtained as a result of processing the {\it Herschel} data in the Hi-GAL project~\citep{2010A&A...518L.100M} in order to determine the abundance of C$_2$H. The processing algorithm was presented by~\cite{2015MNRAS.454.4282M} and was applied to the objects from the Hi-GAL survey by \citet{2017MNRAS.471.2730M}. The data were taken from the open sources\footnote{http://www.astro.cardiff.ac.uk/research/ViaLactea/}. Angular resolution of the $N_{\rm H_2}$ and $T_{\rm dust}$ maps was 12\arcsec.

\section{Analysis}\label{section3}

To estimate the C$_2$H column density ($N_{\rm C_2H}$, cm$^{-2}$), we checked that the ratio between the hyperfine components brightness temperatures is consistent with optically thin emission. The ratio of the components at 87328.624~MHz to 87316.925~MHz was in the range of 0.48--0.52 (close to the theoretical 0.50) in the observed positions. We also tried to analyse the intensity ratio of the multiplet using the CLASS {\it hfs} procedure (that assumes that the widths of all lines in the multiplet are equal). The approximation did not converge due to the asymmetric line profiles (see below).

To measure $N_{\rm C_2H}$, we applied an LTE approximation in the optically thin case (Eq.~\ref{formula1}). Also, we used the Rayleigh-Jeans relation $h\nu~\ll~kT$, which is valid in observations at 3~mm. The background radiation temperature ($T_{\rm bg}=2.73$~K) was not taken into account, since its value is negligible. Thus, the column density is determined as follows \citep[see, e.g.,][]{Mangum2015}:

\begin{equation}
    \label{formula1}
    N_{\rm C_2H}={8\pi\nu_{\rm ul}^2{\rm { k}} \over A_{\rm ul} {\rm { h}} c^3} \frac{Q}{g_{\rm u}} \frac{1}{f}\exp{ \left({E_{\rm u}\over k T_{\rm ex}} \right) } W_{\rm C_2H},
\end{equation}
where $\nu_{\rm ul}$ is the average weighted frequency of transition u~$\to$~l~(Hz), that is computed taking into account the line strength of each hyperfine component (see Table~\ref{table3}), $A_{\rm ul}$ is the Einstein coefficient for spontaneous emission (s$^{-1}$), $g_{\rm u}$ is the statistical weight of the level,  $Q$ is the rotational partition function, $E_{\rm u}$ is the the energy of the upper level (ergs),  k is the Boltzmann constant (ergs~K$^{-1}$), h is the Planck constant (ergs~s), c is the speed of light (cm~s$^{-1}$), $T_{\rm ex}$ is the excitation temperature (K), $W_{\rm C_2H}$ is the integrated intensity of the spectral line (K~km~s$^{-1}$), $f$ is the filling  factor of the telescope beam.

The Einstein coefficient for entire transition $A_{\rm ul}$ is, in general form, given by the formula:

\begin{equation}
    \label{formula2}
    A_{\rm ul}={64\pi^4 \over 3{\rm hc ^3}}\nu^3_{\rm ul}{S_{\rm ul} \over g_{\rm u}}\mu^2,
\end{equation}
where the strength of the transition $S_{\rm ul}=1$ in our case because we compute column density using the integrated intensity of the entire multiplet; $\mu=7.7\times 10^{-19}$~CGS units (0.77~D) is the dipole moment of the molecule. 

Substituting Eq.~\ref{formula2} into Eq.~\ref{formula1}, we obtain the following equation to estimate the column density:

\begin{equation}
    \label{formula4}
    N_{\rm C_2H}={3 {\rm k} \over 8 \pi^3 \nu_{\rm ul} \mu^2} \frac{Q}{f} \exp{ \left({E_{\rm u}\over k T_{\rm ex}} \right) } W_{\rm C_2H}.
\end{equation} 

\begin{table}
	\centering
	\caption{Parameters of the C$_2$H(1--0) hyperfine structure taken from CDMS.}
	\label{table3}
	\begin{tabular}{ccccc}
		\hline
		Transition & Frequency &  $E_{\rm u}/{\rm k}$ & $g_{\rm u}$ & $S_{\rm ul}$\\
		        & [MHz]      & [K] &   & \\
		\hline
		 1$_{3/2, 1}$ -- 0$_{1/2, 1}$ & 87284.156 & 4.2 & 3 & 0.17 \\
		 1$_{3/2, 2}$ -- 0$_{1/2, 1}$ & 87316.925 & 4.2 & 5 & 1.66\\
		 1$_{3/2, 1}$ -- 0$_{1/2, 0}$ & 87328.624 & 4.2 & 3 & 0.83\\
		 1$_{1/2, 1}$ -- 0$_{1/2, 1}$ & 87402.004 & 4.2 & 3 & 0.83\\
		 1$_{1/2, 0}$ -- 0$_{1/2, 1}$ & 87407.165 & 4.2 & 1 & 0.33\\
		 1$_{1/2, 1}$ -- 0$_{1/2, 0}$ & 87446.512 & 4.2 & 3 & 0.17\\
		\hline
    \end{tabular}
\end{table}

All the parameters in Eq.~\ref{formula4}, excluding the product $Q \exp{({E_{\rm u} / ({\rm k} T_{\rm ex})})}  W_{\rm C_2H}$, only depend on the transition type and the structure of a molecule. The value of $Q$ depends on the excitation temperature $T_{\rm ex}$, while $W_{\rm C_2H}$ is determined from observations. For each observed position we computed $Q$ from the adopted value of $T_{\rm ex}$, using an expression for linear molecules:

\begin{equation}
    \label{formula5}
    Q=\frac{{\rm { k}}~T_{\rm ex}}{{\rm { h}}~B_0} + {1 \over 3},
\end{equation}
where $B_0~=~43674.534~\times~10^6$~s$^{-1}$ is the rotational constant for the molecule.

We used the dust temperature $T_{\rm dust}$ as the excitation temperature $T_{\rm ex}$. To do this, in the dust temperature map we selected the regions with a diameter of 28\arcsec{} (our beam size), centered at the positions of our observations (see Fig.~\ref{fig:DSS}) and averaged the $T_{\rm dust}$ value over our beam size. The standard deviations from the mean value was used as the uncertainty of the $T_{\rm dust}$ and $N_{\rm H_2}$ values. Below we discuss a possible inconsistency between the dust temperature and the excitation temperature for the C$_2$H(1--0) transition.

Molecular abundance ($x_{\rm C_2H}$) was determined relative to the column density of the hydrogen nuclei (Eq.~\ref{formula7}):

\begin{equation}
    \label{formula7}
    x_{\rm C_2H}= {N_{\rm C_2H} \over 2N_{\rm H_2}}.
\end{equation}

The values of integrated intensity $W_{\rm C_2H}$ were determined as an integral of the brightness temperature $T_{\rm mb}$ over velocity of the emission spectra using CLASS.

\section{Results}\label{section4}

\begin{figure*}
\includegraphics[width=15cm]{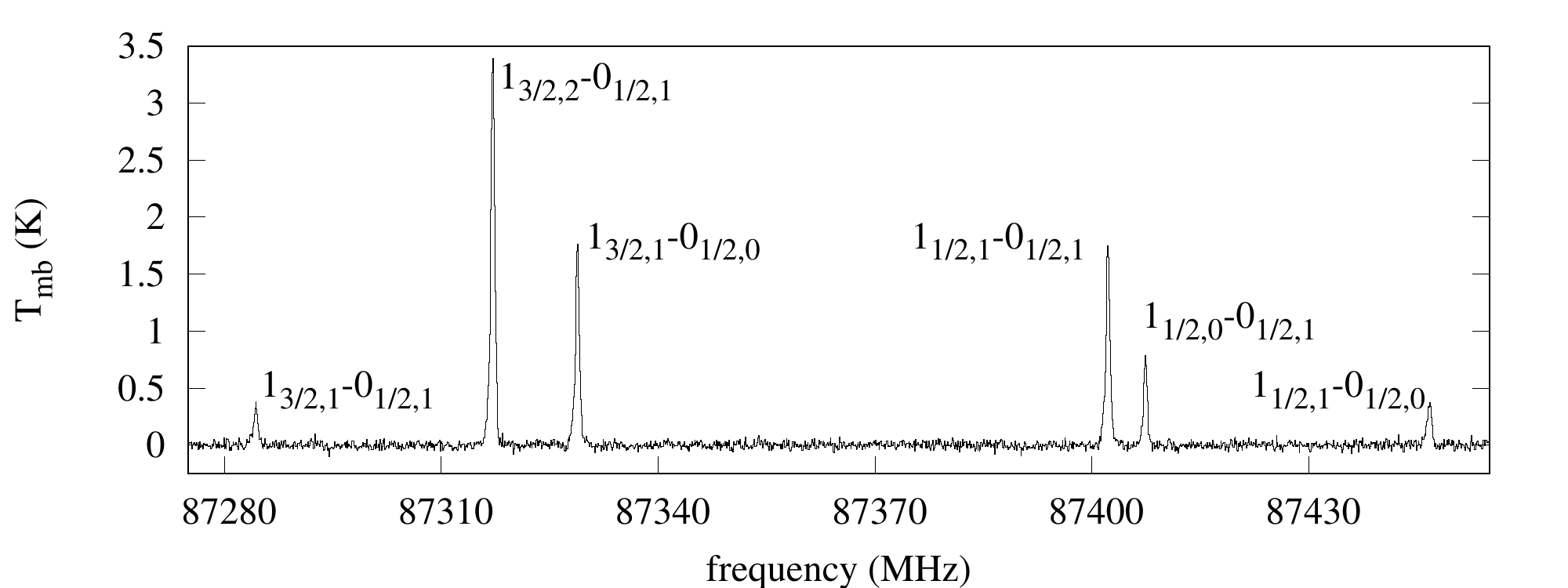}
\caption{Emission spectrum of the C$_2$H(1--0) transition at the position~10.}\label{fig:pos10}
\end{figure*}

\begin{figure*}
\includegraphics[width=15cm]{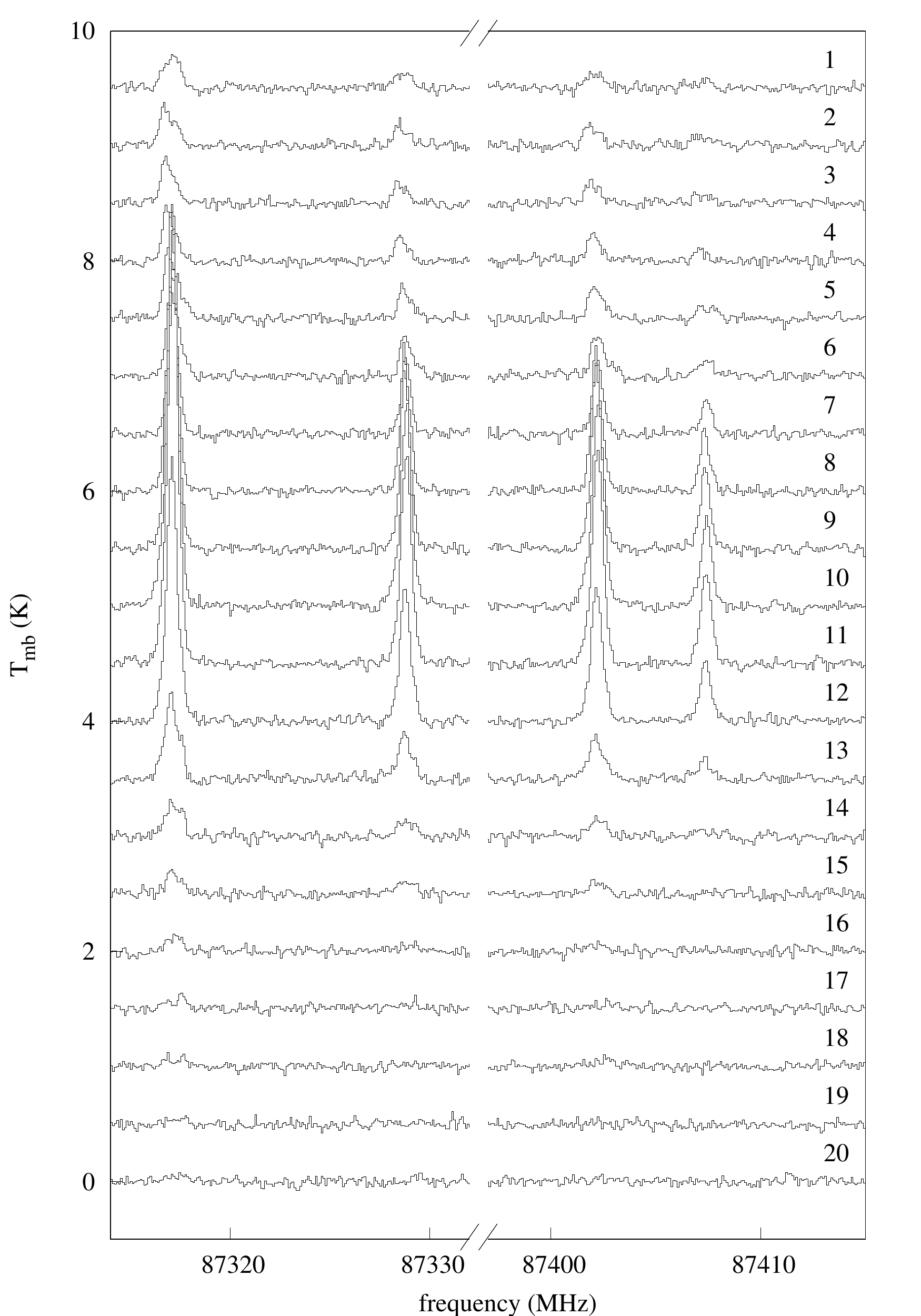}
\caption{The brightest hyperfine components of the C$_2$H(1--0) transition. The position numbers correspond to Fig.~\ref{fig:DSS}.}\label{fig:spectra}
\end{figure*}

The hyperfine components of the ethynyl $J=1-0$ transition were detected at all observed positions. The spectrum of the entire hyperfine structure at position~10 is shown in Fig.~\ref{fig:pos10}, while the four brightest components at all positions are shown in Fig.~\ref{fig:spectra}. The brightest lines were found at positions 8--12, towards the molecular cloud. The hyperfine components (at frequencies $\nu=87284.156$~MHz and $\nu=87446.512$~MHz) are poorly detected towards the ionizing star LS~19 (position~1). Towards HD~253327 (position~18), only the brightest components of the hyperfine structure were detected. The parameters of the ethynyl emission lines are listed in Table~\ref{tab:specparam}. The line profiles are asymmetric, they differ from the Gaussian shape by a higher intensity on the red side (see, for example, the spectra at positions 2--4 and 13, close to the ionizing stars). The flattened line profiles are also observed (for instance, positions 3 and 14). The line profiles of the hyperfine components can be different at the same position. The line profiles are symmetric at the central positions 8--12. The line widths towards the ionizing stars ($\Delta v=3.5\pm 0.17$~\kms{} at position~1; $\Delta v=5.12\pm0.71$~\kms{} at position 18) are 1.5–2 times higher than the widths towards the molecular cloud ($\Delta v=2.37\pm0.17$~\kms{} at position 10). The velocity of the line peak towards the molecular cloud ($v=7.31$~\kms) corresponds to the dip of the double-peaked line towards the ionizing stars (for example, at positions 2, 3, 5, 14, 17, and 18, peaks at the velocities $\approx 5.5$ and 8.2~\kms). The distance between the peaks at the positions with the double-peaked line is approximately 1--1.5~\kms. Such a difference in velocities corresponds to the characteristic expansion rate of the spherically-symmetric \hii{} regions \citep[see, e.g.,][]{2009ARep...53..611K, 2019MNRAS.488.5641K}. All this indicates the presence of several kinematic components along the line of sight near the ionizing stars. Apparently, we observe the front and back molecular walls of the \hii{} regions, which contribute to the formation of a blended components with close velocities. Examples of position~-- velocity diagrams for expanding envelopes around star-forming complexes with several \hii{} regions can be found in works by \citet{2003AstL...29...77L, 2020AJ....160...66P}. Despite the quantitative difference between the characteristic expansion rates of star-forming complexes and \hii{} regions around single stars, the qualitative picture is similar \citep[see also][]{2020MNRAS.497.2651K}.

\begin{table}
	\centering
	\caption{The parameters of the C$_2$H(1--0) multiplet. The parameters marked by an asterisk (*) are determined using the {\it hfs} method, which considers them to be equivalent to a single Gaussian with different levels of relative intensity. The peak intensity $T_{\rm peak}$ was measured for the brightest component at the frequency $\nu=87316.925$~MHz using the {\it gauss} method.}
	\label{tab:specparam}
	\begin{tabular}{ccccc} 
		\hline
		Position & $v^{*}$, & $\Delta v^{*}$, & $T_{\rm peak}$,  & $W_{\rm C_2H}$, \\
		        & [\kms]  & [\kms]       & [K]               & [K~\kms] \\
		\hline
		1 & 7.59 $\pm$ 0.17 & 3.43 $\pm$ 0.17 & 0.33 $\pm$ 0.02 & 2.98 $\pm$ 0.36  \\
		2 & 8.24 $\pm$ 0.17 & 3.08 $\pm$ 0.17 & 0.36 $\pm$ 0.03 & 3.45 $\pm$ 0.39  \\
		3 & 8.24 $\pm$ 0.17 & 2.90 $\pm$ 0.17 & 0.42 $\pm$ 0.03 & 3.32 $\pm$ 0.38  \\
		4 & 8.06 $\pm$ 0.17 & 2.63 $\pm$ 0.17 & 0.50 $\pm$ 0.03 & 3.47 $\pm$ 0.33  \\
		5 & 7.59 $\pm$ 0.17 & 3.07 $\pm$ 0.17 & 0.62 $\pm$ 0.04 & 4.89 $\pm$ 0.50  \\
		6 & 7.28 $\pm$ 0.17 & 2.73 $\pm$ 0.17 & 0.81 $\pm$ 0.04 & 5.68 $\pm$ 0.51  \\
		7 & 7.30 $\pm$ 0.17 & 2.09 $\pm$ 0.74 & 1.49 $\pm$ 0.03 & 8.18 $\pm$ 0.65  \\
		8 & 7.47 $\pm$ 0.17 & 2.12 $\pm$ 0.17 & 2.63 $\pm$ 0.06 & 15.09 $\pm$ 1.16  \\
		9 & 7.45 $\pm$ 0.17 & 2.19 $\pm$ 0.17 & 3.29 $\pm$ 0.06 & 19.80 $\pm$ 0.99  \\
		10 & 7.31 $\pm$ 0.17 & 2.37 $\pm$ 0.17 & 3.66 $\pm$ 0.11 & 24.46 $\pm$ 1.85  \\
		11 & 7.32 $\pm$ 0.17 & 2.33 $\pm$ 0.17 & 3.86 $\pm$ 0.08 & 25.23 $\pm$ 1.81 \\
		12 & 7.46 $\pm$ 0.17 & 2.41 $\pm$ 0.17 & 2.45 $\pm$ 0.07 & 16.95 $\pm$ 1.26 \\
		13 & 7.55 $\pm$ 0.17 & 3.02 $\pm$ 0.17 & 0.77 $\pm$ 0.04 & 6.09 $\pm$ 0.52 \\
		14 & 7.21 $\pm$ 0.17  & 3.47 $\pm$ 0.17 & 0.33 $\pm$ 0.02 & 2.82 $\pm$ 0.37 \\
		15 & 7.25 $\pm$ 0.17 & 3.33 $\pm$ 0.19 & 0.21 $\pm$ 0.02 & 2.16 $\pm$ 0.32  \\
		16 & 7.14 $\pm$ 0.17 & 3.77 $\pm$ 0.34  & 0.16 $\pm$ 0.01 & 1.49 $\pm$ 0.43  \\
		17 & 6.52 $\pm$ 0.23 & 4.27 $\pm$ 0.58  & 0.10 $\pm$ 0.01 & 1.43 $\pm$ 0.43 \\
		18 & 6.50 $\pm$ 0.29 & 5.12 $\pm$ 0.71  & 0.07 $\pm$ 0.01 & 1.02 $\pm$ 0.30 \\
		19 & 6.57 $\pm$ 0.50 & 4.12 $\pm$ 1.01  & 0.06 $\pm$ 0.01 & 0.24 $\pm$ 0.25   \\
		20 & 5.93 $\pm$ 0.31 & 4.11 $\pm$ 0.72  & 0.06 $\pm$ 0.02 & 0.44 $\pm$ 0.22 \\
		\hline
    \end{tabular}
\end{table}

\begin{figure*}
\centering
\includegraphics[width=0.45\linewidth]{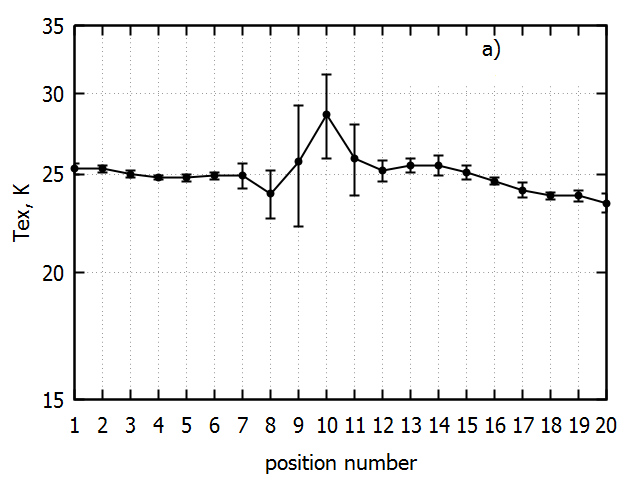}
\includegraphics[width=0.45\linewidth]{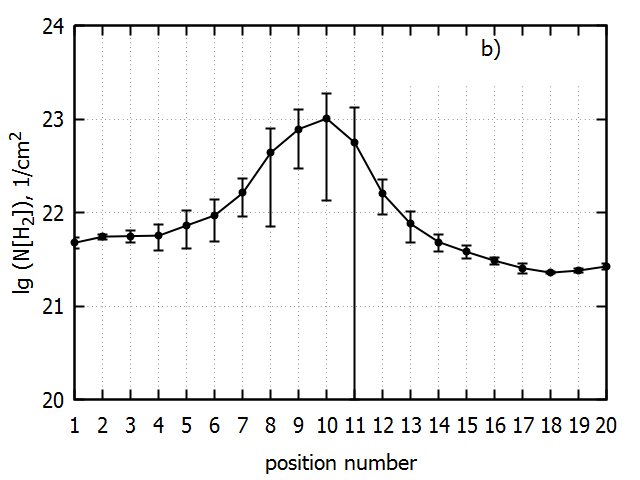}\\
\includegraphics[width=0.45\linewidth]{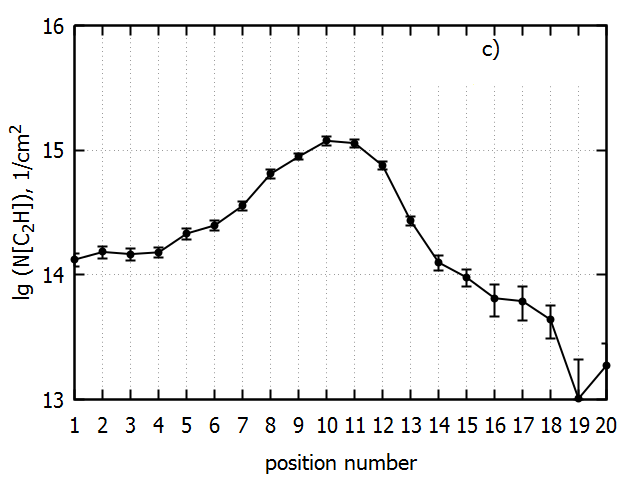}
\includegraphics[width=0.45\linewidth]{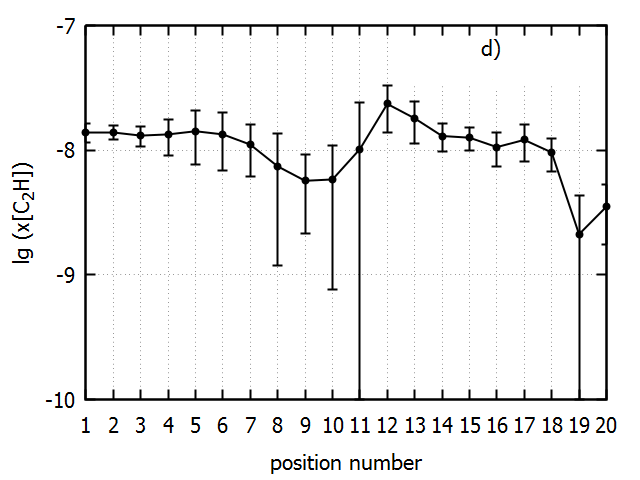}
\caption{The profiles of the physical parameters: a) dust temperature adopted as the $T_{\rm ex}$ value, b) the H$_2$ column density, c) the column density of C$_2$H; d) the abundance of ethynyl.}
\label{fig:results}
\end{figure*}

The values of $T_{\rm ex}$ are listed in Table~\ref{tab:coldens} and shown in Fig.~\ref{fig:results}a. The excitation temperature reaches its maximum of $T_{\rm ex}=(28.6 \pm 2.7)$~K at the central position~10 (towards S255~IRS1 in the molecular cloud) and then decreases at most by 5~K towards the ionizing stars. $N_{\rm H_2}$ reaches its maximum towards S255~IRS1. Radial profile of $N_{\rm H_2}$ is shown in Fig.~\ref{fig:results}b. It is similar to that of $T_{\rm ex}$ with a maximum at the position 10 and a gradual decrease towards both ionizing stars. The fact that $N_{\rm H_2}$ does not decrease to zero at the positions 1 and 18 indicates the presence of neutral gas towards the ionizing stars. This is consistent with the assumption about the front and back walls of the \hii{} regions.

The column densities and the abundances of ethynyl are presented in Table~\ref{tab:coldens} and shown in the lower panels in Fig.~\ref{fig:results}. The radial profile of $N_{\rm C_2H}$ has the same shape as that of $N_{\rm H_2}$. The maximum column density  $N_{\rm C_2H} = 11.78 \pm 0.91 \times 10^{14}$~cm$^{-2}$ is reached at the position~10. At the positions 1 and 18, the column density of ethynyl decreases by an order of magnitude in S255 and 25 times in S257.

Having determined the value of $N_{\rm C_2H}$ under the LTE assumption, we tested how close are the LTE and non-LTE estimates. For that purpose, we used the RADEX package, which allows estimating the brightness of the emission lines in the non-LTE model for given parameters of the medium: temperature, density, line width, and column density of molecules \citep[][]{2007A&A...468..627V}. We adopted the dust temperature as the kinetic temperature, the observed line widths and our LTE-based estimates for $N_{\rm C_2H}$. Since we have no information on the gas density in the molecular cloud, we estimated the brightness of the theoretical spectra for several density values, namely 10$^2$, 10$^3$, and 10$^4$~cm$^{-3}$. For the central positions, we achieved the agreement of the observed brightness with the LTE and non-LTE models within 10\% of the intensities for the density of 10$^4$ cm$^{-3}$. This value of density is reasonable for molecular clouds. For the directions to the ionizing stars, the theoretical non-LTE values of the brightness temperatures calculated for the densities of 10$^2$-10$^3$~cm$^{-3}$ (reasonable values of the density when moving from molecular cloud to a star) are significantly (up to an order of magnitude) lower than the observed. The calculated excitation temperatures are lower than the adopted kinetic temperatures in those directions. To obtain the observed line brightness, we would need to increase $N_{\rm C_2H}$ by an order of magnitude. On the other hand, in the non-LTE simulation with the density of 10$^4$~cm$^{-3}$, brightness of hyperfine components is reproduced within 10\% towards the ionizing stars also, which might be related to the presence of the dense front and back molecular walls of the \hii{} regions. The $T_{\rm ex}$ values, obtained in the non-LTE case, are of 4-6~K. Therefore, the ethynyl lines are excited in sub-thermal regime. As the $N_{\rm C_2H}$ weakly depend on excitation temperature at 3~mm, varying $T_{\rm ex}$ from 4-6 K to 25~K, we obtain the results that differ within a factor of 2. To diminish the range of parameters in the non-LTE modelling, it is necessary to carry out additional observations of ethynyl lines corresponding to other transitions.

The ethynyl abundances at the positions 1-8 (towards the LS~19 star and around, in S255) and 13--18 (HD~253327 and around, in S257) are higher in comparison with the values at the positions 9-12. For example, towards the molecular cloud, at the position 10, the abundance is $x_{\rm C_2H}=5.83\pm5.07 \times 10^{-9}$, while towards the ionizing stars (positions 1 and 18) $x_{\rm C_2H}=13.91\pm 2.47 \times 10^{-9}$ and $x_{\rm C_2H}=9.60\pm2.83 \times 10^{-9}$. A significant uncertainty in determination of the ethynyl abundance towards the molecular cloud (for example, $x_{\rm C_2H}=10.18\pm13.81\times 10^{-9}$ at the position 11)  is due to the fact that at the positions over which the value of $N_{\rm H_2}$ is averaged, there are several bright spots in the $N_{\rm H_2}$ map. For this reason, the standard deviation at the positions 8-11 is higher than at other positions. Therefore, these uncertainties are formal values, and the real uncertainties are determined by the brightness of the ethynyl lines. Nevertheless, we clearly see the factor of 2 decrease in the $x_{\rm C_2H}$ values towards the molecular cloud in comparison with rather flat areas of the radial profile towards the \hii{} regions. The flat parts of the abundance profile suggest that the \hii{} regions are surrounded by quasi-spherical molecular shells in which ethynyl is distributed almost uniformly.

\begin{table}
	\centering
	\caption{The column densities and abundances of ethynyl in S255 and S257.}
	\label{tab:coldens}
	\begin{tabular}{ccccc} 
		\hline
		Position &  $T_{\rm ex}$, & $N_{\rm H_2}$,   & $N_{\rm C_2H}$,       & $x_{\rm C_2H}$, \\
		        &              [K]          & $10^{20}$~[cm$^{-2}$] & $10^{14}$~[cm$^{-2}$] & $10^{-9}$    \\
		\hline
		1  & 25.3 $\pm$ 0.3 & 47.6 $\pm$ 6.2 & 1.33 $\pm$ 0.16 & 13.9 $\pm$ 2.5\\
		2  & 25.3 $\pm$ 0.2 & 55.3 $\pm$ 3.6 & 1.53 $\pm$ 0.17 & 13.9 $\pm$ 1.8\\
    	3  & 25.0 $\pm$ 0.2 & 56.1 $\pm$ 7.6 & 1.47 $\pm$ 0.17 & 13.1 $\pm$ 2.3\\
		4  & 24.8 $\pm$ 0.1 & 56.7 $\pm$ 17.4 & 1.52 $\pm$ 0.14 & 13.4 $\pm$ 4.3\\
		5  & 24.8 $\pm$ 0.2 & 72.9 $\pm$ 31.2 & 2.15 $\pm$ 0.22 & 14.7 $\pm$ 6.5\\
		6  & 24.9 $\pm$ 0.2 & 93.5 $\pm$ 44.7 & 2.50 $\pm$ 0.22 & 13.4 $\pm$ 6.5\\
		7  & 24.9 $\pm$ 0.7 & 162.2 $\pm$ 71.2 & 3.60 $\pm$ 0.29 & 11.1 $\pm$ 4.9\\
		8  & 23.9 $\pm$ 1.3 & 436.6 $\pm$ 365.2 & 6.48 $\pm$ 0.50 & 7.4 $\pm$ 6.2 \\
		9  & 25.7 $\pm$ 3.5 & 780.9 $\pm$ 483.2 & 8.89 $\pm$ 0.49 & 5.7 $\pm$ 3.5\\
		10 & 28.6 $\pm$ 2.7 & 1011.0 $\pm$ 877.0 & 11.78 $\pm$ 0.91 & 5.8 $\pm$ 5.1\\
		11 & 25.9 $\pm$ 2.1  & 559.3 $\pm$ 758.1 & 11.38 $\pm$ 0.83 & 10.2 $\pm$ 13.8\\
		12 & 25.2 $\pm$ 0.6  & 159.9 $\pm$ 65.1 & 7.52 $\pm$ 0.56 & 23.5 $\pm$ 9.7\\
		13 & 25.5 $\pm$ 0.4 & 75.7 $\pm$ 27.7 & 2.72 $\pm$ 0.23 & 18.0 $\pm$ 6.8\\
		14 & 25.5 $\pm$ 0.6 & 48.5 $\pm$ 10.5 & 1.26 $\pm$ 0.17 & 13.09 $\pm$ 3.3\\
		15 & 25.1 $\pm$ 0.4 & 38.0 $\pm$ 6.0 & 0.96 $\pm$ 0.14 & 12.6 $\pm$ 2.7\\
		16 & 24.6 $\pm$ 0.2 & 30.6 $\pm$ 2.4 & 0.65 $\pm$ 0.19 & 10.6 $\pm$ 3.2\\
		17 & 24.1 $\pm$ 0.4 & 25.5 $\pm$ 3.0 & 0.62 $\pm$ 0.19 & 12.1 $\pm$ 3.9\\
		18 & 23.8 $\pm$ 0.2 & 22.8 $\pm$ 0.6 & 0.44 $\pm$ 0.13 & 9.6 $\pm$ 2.8\\
		19 & 23.8 $\pm$ 0.3 & 24.0 $\pm$ 1.4 & 0.10 $\pm$ 0.11 & 2.1 $\pm$ 2.2\\
		20 & 23.4 $\pm$ 0.5  & 26.6 $\pm$ 1.7 & 0.19 $\pm$ 0.09 & 3.5 $\pm$ 1.8\\
		\hline
    \end{tabular}
\end{table}

\section{Discussion}\label{discussion}

For the efficient formation of ethynyl, it is necessary to have C$^+$ ion and molecular hydrogen H$_2$ \citep[see][]{1976RPPh...39..573D, 1995ApJS...99..565S, 2013ChRv..113.8981S} in the gas-phase; these conditions are met in PDRs. In molecular clouds, carbon is bounded mainly in CO molecules due to the shielding of stellar UV~radiation by dust and H$_2$ molecules. Therefore, the $x_{\rm C_2H}$ value decreases in the molecular cloud at the positions 7-13 due to the conversion of carbon into CO. Approximately equal abundances of ethynyl towards the \hii{} regions indicates that the chemical evolution of the surrounding PDRs proceeds in a similar way. This is not surprising, since these \hii{} regions have similar sizes and they are ionised by stars of similar spectral types. We note that several point sources are visible in the optical (e.g.,~Fig.~\ref{fig:DSS}) and IR~images (e.g.,~2MASS) at the positions 12-13. In particular, the position 12 contains an emission-line star S255~1, with observed H$\alpha$~line~\citep[][bright spots are also visible in our Fig.~\ref{fig:DSS}]{2005A&A...440..569A} and a Herbig-Haro object S255~H$_2$~1~\citep{1997ApJ...488..749M}. Nearby X-ray source CXOU~J061250.6+175909, a young low-mass star in the S255-IR cluster, is located there \citep{2011A&A...533A.121M}. It is quite possible that the ionizing radiation in the vicinity of these point sources leads to a local increase of the ethynyl abundance due to the ionization of carbon-containing molecules with the release of the C$^+$ ion, which takes part in the chemical chains of the ethynyl formation. There are no specific features in the radial profiles of the ethynyl abundances towards the ionization front (positions 6 and 16 for S255 and S257, respectively). The values of $T_{\rm dust} \approx 23-25$~K at these positions correspond to the UV~field with $G_0=50-100$ in units of the Habing field \citep[see, for example][]{2020MNRAS.497.1050K}. The values of $G_0$ in S255--S257 are close to that found in the Horsehead~PDR, therefore it is necessary to make additional observations of other ethynyl transitions in order to obtain accurate radial profiles of the abundances and to compare ethynyl abundances at the ionised edge of the Horsehead PDR and in the vicinity of the ionization front in S255--S257.

\citet{2020ApJ...889...43Z} observed the C$_2$H~(4--3) lines towards S255~IRS1 with ALMA and estimated the ethynyl column density there. Despite the difference in the spatial resolution, the estimates agree within a factor of 5 (our estimate is lower), indicating that ethynyl is present in the dense molecular cloud. 

The values of the $N_{\rm C_2H}$ in the vicinity of the ionizing stars are similar to those found in the Horsehead, IC~63~\citep{2004A&A...417..135T}, and M~8~\citep{2019A&A...626A..28T}. Moving closer to the gas column density peak in the molecular cloud, we find higher values of $N_{\rm C_2H}$ similar to the values observed in the Orion~Bar~PDR \citep{2015A&A...575A..82C} and some other molecular clouds \citep[see, e.g.,][]{2016ARep...60..904P, 2020A&A...636A..19B}. The maximum observed values of $N_{\rm C_2H}$ at the positions 10 and 11 are comparable with the values found in OMC-1 \citep[][see also Fig.~10 in \citet{2020A&A...636A..19B}]{1997ApJ...482..245U}. The abundances of ethynyl in S255--257 are similar to the value observed in the Orion Bar PDR. Namely, according to the estimates made by \citet{2015A&A...575A..82C}, the abundance of $(0.7-2.7)\times 10^{-8}$ in an object with the radiation field $G_0 \approx 10^4-10^5$ can be explained using the high-temperature chemical reactions with excited molecular hydrogen. In S255--257, a comparable abundance of ethynyl is observed at by two orders of magnitude lower $G_0$. Thus, the PDRs in S255--S257 are interesting for astrochemical modelling.

\section{Conclusions}\label{conclusions}

We summarise our main results below:

\begin{enumerate}
    \item We present observations of the C$_2$H molecule in the \hii{} regions S255 and S257 at twenty positions along the straight lines connecting the IR~source S255~IRS1, located in the molecular cloud, with the ionizing stars in each of the regions, LS~19 and HD~253327.
    
    \item The brightest emission of ethynyl is observed towards the molecular cloud (brightness temperature $\approx 4$~K), the weakest one – towards the ionizing stars ($<0.5$~K). The largest line widths are found towards the ionizing stars ($\approx 3-5$~\kms), while the smallest ones are found towards the molecular cloud in between the \hii{} regions ($\approx 2$~\kms). The spectral line profiles indicate that several kinematic components exist in the molecular envelopes of the \hii{} regions. The components are resolved on the line of sight towards the ionizing stars. The components can be the front and back walls of the \hii{} regions.
    
    \item  We determine the column densities and the abundances of ethynyl at all twenty studied positions. The maximum value of the column density $\approx 12 \times 10^{14}$~cm$^{–2}$ is found towards the centre of the molecular cloud; it decreases closer to the ionizing stars in both directions. The minimum abundances, on the contrary, are found towards the molecular cloud ($6 \times 10^{-9}$), while towards the ionizing stars the abundances are about two times higher. The maximum ethynyl abundance of $2 \times 10^{-8}$ is found towards the point sources in the molecular cloud, namely, emission-line stars or X-ray sources.
\end{enumerate}

\section*{Acknowledgements}

We are grateful to L.~E.~Pirogov, S.~V.~Kalenskii for valuable advice on the observational data processing, to D.~A.~Semenov for the discussion on the chemistry of ethynyl, and to the anonymous referee for valuable comments.

This study was supported by Russian Foundation for Basic research (contract No. 20-02-00643 A). A.~F. Punanova is a member of the Max-Planck-Gesellschaft partner group in the Ural Federal University. A.~F. Punanova is supported by the Russian Ministry of Science and Higher Education via state assignment FEUZ-2020-0038.

\bibliographystyle{mnras}
\bibliography{main} 

\bsp	
\label{lastpage}
\end{document}